\documentclass[prb,twocolumn,showpacs,preprintnumbers,amsmath,amssymb]{revtex4}
\usepackage{graphicx}
\usepackage{dcolumn}
\usepackage{bm}

\preprint{submitted to PRB}

\begin{document}

\title{Origin and Properties of the Gap in 
the Half-Ferromagnetic Heusler Alloys}   

\author{I. Galanakis}\email{I.Galanakis@fz-juelich.de}
\author{P. H. Dederichs}
\affiliation{Institut f\"ur Festk\"orperforschung, Forschungszentrum J\"ulich, 
D-52425 J\"ulich, Germany}
\author{N. Papanikolaou}
\affiliation{Fachbereich Physik, Martin-Luther Universit\"at, Halle-Wittenberg,
D-06099 Halle, Germany}

\date{\today}

\begin{abstract}
We study the origin of the gap and the role of chemical composition 
in the half-ferromagnetic  Heusler alloys using the full-potential screened 
KKR method. In the paramagnetic phase the $C1_b$ compounds, like NiMnSb, 
present a gap. Systems with 18 valence electrons, $Z_t$, per unit cell, like CoTiSb,
are semiconductors, but when $Z_t > 18 $ antibonding states are also 
populated, thus the paramagnetic phase becomes unstable and the half-ferromagnetic
one is stabilized. The minority  occupied bands  accommodate a total of nine electrons 
and the total magnetic moment per unit cell in $\mu_B$ is just the difference between 
$Z_t$  and $2\times 9$. While the substitution of the transition metal atoms 
may preserve the half-ferromagnetic character, substituting the $sp$ atom
results in a practically rigid shift of the bands and the loss of half-metallicity.
Finally we show that expanding or contracting the lattice parameter by 
2\%\ preserves the  minority-spin gap.
\end{abstract}


\maketitle

\section{Introduction\label{sect1}}
             
Heusler alloys\cite{heusler} have attracted during the last century a great interest due
to the possibility to study in the same family of alloys a series of interesting diverse 
magnetic phenomena like itinerant and localized magnetism, antiferromagnetism,
helimagnetism, Pauli paramagnetism or heavy-fermionic 
behavior.\cite{landolt,Pierre97,Tobola00} 
The first Heusler alloys studied were crystallizing 
in the $L2_1$ structure which consists of 4 fcc sublattices. Afterwards, it was
discovered that it is possible to substitute one of the four sublattices with 
vacancies ($C1_b$ structure). The latter compounds are often called 
half-Heusler alloys, while the $L2_1$ compounds are often referred to as 
full-Heusler alloys. In 1983 de Groot and  his collaborators\cite{groot} 
showed that one of the half-Heusler compounds, NiMnSb, is a half-ferromagnet, 
i.e. the minority band is semiconducting with a gap at the Fermi level $E_F$, 
leading to 100\% spin polarization at $E_F$.
Recently the rapid development of magnetoelectronics intensified the interest on such 
materials. Adding the spin degree of freedom to the conventional electronic 
devices has several advantages like the nonvolatility, the increased data processing
speed, the decreased electric power 
consumption and the increased integration densities.\cite{Wolf} The current advances in
new materials are promising for engineering new spintronic devices in the near 
future.\cite{Wolf}  
Other known half-ferromagnetic materials are CrO$_2$,\cite{Soulen98} 
La$_{0.7}$Sr$_{0.3}$MnO$_3$,\cite{Soulen98} the diluted 
magnetic semiconductors\cite{Akai98} like (In,Mn)As and very recently also CrAs 
in the zinc-blende structure was proposed to be a half-ferromagnet.\cite{Akinaga00}
Although thin films of CrO$_2$ and La$_{0.7}$Sr$_{0.3}$MnO$_3$  
have been verified to present practically
100\% spin-polarization at the Fermi level at low temperatures,\cite{Soulen98,Park98} 
NiMnSb  remains attractive for technical applications like 
spin-injection devices,\cite{Data}  spin-filters,\cite{Kilian00} 
tunnel junctions,\cite{Tanaka99} or GMR devices\cite{Caballero98}
due to its relatively  high  Curie temperature ($T_c\simeq$730K) compared to 
these compounds.\cite{landolt}

The half-ferromagnetic character of NiMnSb in single crystals has been 
well-established experimentally. Infrared 
absorption\cite{Kirillova95} and spin-polarized positron-annihilation\cite{Hanssen90}
gave a spin-polarization of $\sim$100\% at the Fermi level.
Recently high quality films of NiMnSb have been grown,\cite{Roy}
 but they were found not to reproduce the half-ferromagnetic character of 
the bulk. Values of 58\% and 50\% for the 
spin-polarization at the Fermi level were obtained 
by Soulen \textit{et al.}\cite{Soulen98} and by
Mancoff \textit{et al.},\cite{Mancoff99} respectively,  and recently
Zhu \textit{et al.}\cite{Zhu01} found a value of 40\% using spin-resolved photoemission
measurements on polycrystalline films.  
Ristoiu \textit{et al.} \cite{Ristoiu00} showed 
that during the growth of the NiMnSb thin films, Sb atoms segregate
to the surface decreasing the 
obtained spin-polarization; they measured a value of 
$\sim$30\% at 200K, while at room temperature the net polarization was 
practically zero. But when they removed the excess of Sb
by a flash annealing, they managed to get a nearly  stoichiometric ordered alloy surface 
terminating in MnSb. Inverse photoemission experiments at room temperature 
revealed that the latter
surface shows a spin-polarization of about 67$\pm$9\% which is significantly 
higher than all previous values.\cite{Ristoiu00}  Finally  there is experimental
evidence that for a temperature of $\sim$80 K there is transition 
from a half metal to a normal ferromagnet,\cite{Hordequin00}
but these experiments are not yet conclusive.

Several groups have verified the half-ferromagnetic character of
bulk NiMnSb using first-principles calculations\cite{iosif,Calculations} and 
its magnetic properties can be well-understood in 
terms of the hybridization between  the higher-valent 3$d$ atom (like Ni) and Mn, and the indirect
exchange of the Mn $d$ electrons through the $sp$ atom.\cite{Reitz79,Kubler83}
Larson \textit{et al.} have shown that the actual structure of NiMnSb is 
the most stable with respect to an interchange of the atoms\cite{Larson00}
and Orgassa \textit{et al.} showed that a few percent of disorder induce states
within the gap but do not destroy the half-metallicity.\cite{Orgassa99}  
Ab-initio calculations\cite{groot2,iosifSURF} have shown that the  NiMnSb surfaces do not
present 100\% spin-polarization at the Fermi level 
and it was proposed by Wijs and de Groot that at some interfaces it is possible to
restore the half-ferromagnetic character of NiMnSb.\cite{groot2}
Jenkins and King\cite{Jenkins01} using a pseudopotential technique showed 
that the MnSb terminated (001) NiMnSb surface relaxes mildly; the Sb atoms 
move slightly outwards and the Mn inwards with a total buckling of only 0.06 \AA. 
They identified two surface states in the minority 
band inside the gap, which cross the Fermi level and 
which are strongly localized in the surface region.

In this contribution we will study the origin of the gap and different ways to
influence the position of the Fermi level that can lead to new 
half-ferromagnetic Heusler materials
which might have advantages as compared to the NiMnSb films and therefore could  be more 
attractive for applications. In Section~\ref{sect2} we present the details of our calculations
and in Section~\ref{sect3} we discuss the properties of the XMnSb compounds, where
X stands for Ni, Co, Rh, Pd, Ir or Pt. In Section~\ref{sect4} we study the influence of the 
lattice parameter on the position of the Fermi level and in Section~\ref{sect5} we investigate the origin 
of the gap.  In Sections~\ref{sect6} and \ref{sect7} we study the influence of
the lower-valent transition metal and of the $sp$ atom, respectively,  on the  
properties of the gap. Finally in Section~\ref{sect8} we conclude and summarize our results.

\begin{figure}
\includegraphics[scale=0.4]{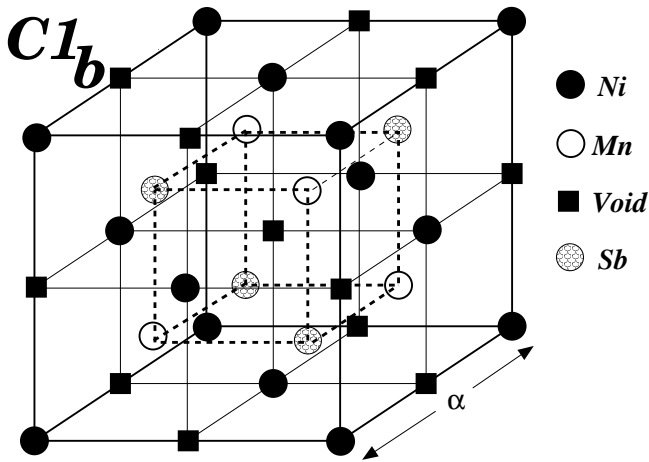}
 \caption{\label{fig1} Schematic representation of the $C1_b$ structure. The lattice
consists of 4 fcc sublattices. The unit cell is that of a fcc lattice with four
atoms as basis: Ni at $(0\:0\:0)$, Mn at $({1\over4}\:{1\over4}\:{1\over4})$, a vacant site at 
$({1\over2}\:{1\over2}\:{1\over2})$  and 
Sb at $({3\over4}\:{3\over4}\:{3\over4})$  in Wyckoff coordinates} 
\end{figure}
  
\section{Computational Details\label{sect2}}

To perform the calculations, we used the Vosko, Wilk and Nusair
parameterization\cite{Vosko} for the local density approximation
(LDA) to the exchange-correlation potential\cite{kohn} to solve
the Kohn-Sham equations within the full-potential screened Korringa-Kohn-Rostoker 
(FKKR) method.\cite{Zeller95,Pap02}  The full-potential is
implemented by using a Voronoi construction  of Wigner-Seitz
polyhedra that fill the space as described in Ref.
\onlinecite{Pap02}. A repulsive muffin-tin potential (4 Ry high) is
used as reference system to screen the free-space long-range
structure constants into exponentially decaying ones.\cite{Zeller97}
  For the screening we took for all compounds
interactions up to the sixth neighbors into account leading to a
tight-binding (TB) cluster around each atom of 65 neighbors. To
calculate the charge density, we integrated along a contour on the
complex energy plane, which extends from the bottom of the band up
to the Fermi level.\cite{Zeller82} Due to the smooth behavior of
the Green's functions for complex energies, only few energy points
are needed; in our calculations we used 42 energy points. For the
Brillouin zone (BZ) integration, special points are used as
proposed by Monkhorst and Pack.\cite{monkhorst} Only few tens of
\textbf{k} are needed to sample the BZ for the complex
energies, except for the energies close to the real axis near the
Fermi level for which a considerably larger number of \textbf{k}-points 
is  needed. We have used a 
30$\times$30$\times$30 {\bf k}-space grid in the full BZ to perform the 
integrations. In addition we used a cut off 
of $\ell_{max}$=6 for
the multipole expansion of the charge density and the potential
and a cut off of $\ell_{max}$=3 for the wavefuctions. Finally in
our calculations the core electrons are allowed to relax during
the self-consistency.

In Fig. \ref{fig1} we show the $C1_b$ structure, which consists of four 
fcc sublattices. The unit cell is that of a fcc lattice with four
atoms as basis at A=$(0\:0\:0)$, B=$({1\over4}\:{1\over4}\:{1\over4})$, C=
$({1\over2}\:{1\over2}\:{1\over2})$ and D=$({3\over4}\:{3\over4}\:{3\over4})$
in Wyckoff coordinates. In the case of NiMnSb the A site is occupied by
Ni, the B site by Mn and the D site by Sb, while the C site is unoccupied. 
The $C1_b$ structure is similar to the $L2_1$  structure adopted by the 
full Heusler alloys, like Ni$_2$MnSb where also the C site is occupied
by a Ni atom.  We should also mention that the zinc-blende structure 
adopted by a large number of semiconductors, like GaAs, ZnSe, InAs etc., 
is also consisting of four fcc sublattices. In the case of GaAs the A 
site is occupied by a Ga atom,
the B site by a As atom, while the C and D sites are empty. This close 
structure similarity makes the 
Heusler alloys compatible with the existing semiconductor technology and thus
very attractive for industrial applications.

\section{X-M\lowercase{n}S\lowercase{b} Compounds with X= C\lowercase{o}, N\lowercase{i}, 
R\lowercase{h}, I\lowercase{r}, P\lowercase{d} and P\lowercase{t} \label{sect3}}

\begin{figure}
\includegraphics[scale=0.35]{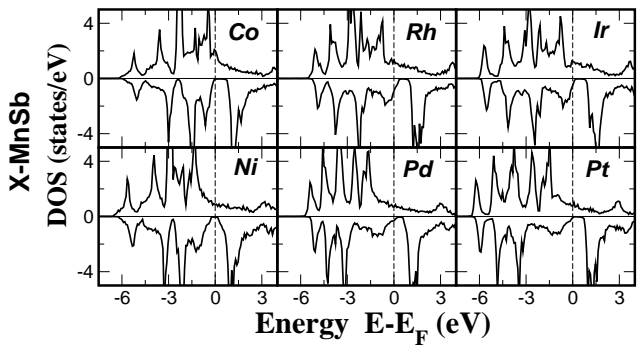}
 \caption{\label{fig2} Spin-projected density of states for the XMnSb 
Heusler alloys.  They all posses a spin-down gap but only in Co-, Ni- and Pt-
based ones is the Fermi level (zero at the energy axis) inside the gap.} 
\end{figure}    

Firstly we calculated the electronic structure of the XMnSb, with X being an 
element of the Co or Ni columns in the periodic table. We used as lattice parameters
the experimental ones for all compounds.\cite{landolt} These 
compounds are known experimentally to be ferromagnets with high Curie temperatures
ranging between 500 K and 700 K for the Co, Ni, Pd and Pt compounds, while 
the Curie temperatures of the Ir and Rh compounds are around room temperature.\cite{landolt}
We should also mention here that IrMnSb is in reality crystallizing in 
the Ir$_{0.92}$Mn$_{1.16}$Sb$_{0.92}$ stoichiometry.\cite{landolt}
The remaining  compounds are known to exhibit a small disorder,\cite{landolt} 
with the exception of CoMnSb.\cite{Otto89} Kaczmarska \textit{et al.} have calculated 
using the KKR method within the coherent potential approximation the spin magnetic
moments for a (Co$_{0.5}$Mn$_{0.5}$)$_2$Sb system, where the Co and Mn sites are 
completely disordered, and found them to agree with the experimental 
values.\cite{Kaczmarska99} In Fig. \ref{fig2} we present
the spin-projected total density of states (DOS) for  all the six compounds. 
We remark that all six compounds present a gap, which      
is more wide in the compounds containing Co, Rh or Ir than 
Ni, Pd or Pt.
Sb $p$ states occupy the lowest part of the DOS shown in the figure, while the Sb $s$ states are
located $\sim$12 eV below the Fermi level.  
For the Ni compound the Fermi level is at the middle of the gap and
for PtMnSb at the left edge of the gap in agreement with previous full-potential 
linear muffin-tin orbitals method (FPLMTO) calculations on these compounds.\cite{iosif}
The gap in the minority NiMnSb band is about 0.5 eV wide in good agreement with the 
experiment of Kirillova and collaborators\cite{Kirillova95} who  analyzing their 
infrared spectra estimated a gap width of $\sim$0.4 eV.
In the case of CoMnSb the gap is considerably larger ($\sim$1 eV) than in
the previous two compounds and the Fermi  level is located at the left 
edge of the spin-down gap. CoMnSb has been studied theoretically by K\"ubler
using the augmented spherical waves (ASW) method. 
He found a DOS similar to ours, with a large gap of 
comparable width and the Fermi level was also located at the left edge 
of the spin-minority gap.\cite{Kubler84} 
For the other three compounds the Fermi level is located below the gap,
although in the case of PdMnSb and IrMnSb it is very near to the edge 
of the gap.

\begin{table}
\caption{\label{table1}Spin magnetic moments in $\mu_B$ using the experimental lattice constants 
(see Ref.~\protect{\onlinecite{landolt}}) except for FeMnSb where we have used 
the lattice parameter estimated in Ref.~~\protect{\onlinecite{Groot86}}. They are calculated
by integrating the spin-projected charge density inside the Wigner-Seitz polyhedron
containing the atom. X stands for the atom occupying the A site (see Fig. \ref{fig1}).}
\begin{ruledtabular}
\begin{tabular}{rrrrrr}
$m^\mathrm{spin}(\mu_B)$ & X & Mn & Sb & Void & Total\\ \hline
NiMnSb & 0.264 & 3.705 & -0.060 & 0.052 & 3.960\\
PdMnSb & 0.080 & 4.010 & -0.110 & 0.037 &4.017 \\ 
PtMnSb & 0.092 & 3.889 & -0.081 & 0.039 &3.938\\
CoMnSb & -0.132 & 3.176 & -0.098 & 0.011& 2.956 \\
RhMnSb & -0.134 & 3.565 & -0.144 & $<$0.001 & 3.287 \\
IrMnSb & -0.192 & 3.332 & -0.114 & -0.003 & 3.022 \\
FeMnSb & -0.702 & 2.715 & -0.053 & 0.019  & 1.979
\end{tabular}
\end{ruledtabular}
\end{table}

The DOS of the different systems   are mainly characterized by
the large exchange-splitting of the Mn $d$ states which is around 3 eV in all cases.
This is clearly seen in the atom-projected DOS of NiMnSb in Fig. \ref{fig4}.
This large exchange splitting leads to large localized spin moments at the 
Mn site; the existence of the localized moments has been verified also 
experimentally.\cite{Yablonskikh01} 
The localization comes from the fact that although $d$ electrons of
Mn are itinerant, the spin-down electrons are almost excluded from the  
Mn site. In Table \ref{table1} we present the 
spin magnetic moments at the different sites for all the compounds under study. 
Here we should mention that in order to calculate the moments we integrate the spin-projected
charge density  inside every Wigner-Seitz polyhedron. 
In our calculations these polyhedra were
the same for every atom in the $C1_b$ structure. Mn moments approach in the case of 
the Ni, Pd and Pt compounds the 4 $\mu_B$ and they agree perfectly with
previous FPLMTO calculations.\cite{iosif} Ni, Pd and Pt 
are ferromagnetically coupled to Mn
with small induced magnetic moments, while in all cases the Sb atom is 
antiferromagnetically coupled to Mn. Overall the calculated moments for the Ni, Pd
and Pt compounds are  in very good agreement with previous \textit{ab-initio}
results.\cite{iosif,Calculations}
Experimental values for the spin-moment at the Mn site can be
deduced from the experiments of Kimura \textit{et al.}\cite{Kimura97} by 
applying the sum rules to their x-ray magnetic circular dichroism 
spectra and the extracted moments agree nicely with our results;
they found a Mn spin moment of 3.85 $\mu_B$ for NiMnSb, 3.95 $\mu_B$ for PdMnSb 
and 4.02 $\mu_B$ for PtMnSb.   
In the case of the Co-, Rh-, and IrMnSb compounds the spin magnetic moment
of the X atom is antiparallel to the Mn localized moment and the Mn moment
is generally about 0.5 $\mu_B$ smaller than in the Ni, Pd and Pt compounds. The Sb atom is here again 
antiferromagnetically coupled to the Mn atom.

\begin{table}
\caption{\label{table2}Number of electrons, $n_{e^-}$, lost (-) or gained (+)
by an atom with respect to the free atom. As was the case for the spin moments 
the total number of electrons is calculated by integrating the total charge 
density inside a polyhedron.}
\begin{ruledtabular}
\begin{tabular}{rrrrr}
$n_{e^-}$ & X & Mn & Sb & Void \\ \hline
NiMnSb & +0.567 & -0.456 & -1.429 & +1.318 \\
PdMnSb & +0.237 & -0.343 & -1.080 & +1.186 \\ 
PtMnSb & +0.034 & -0.212 & -1.041 & +1.219 \\
CoMnSb & +0.516 & -0.411 & -1.426 & +1.321 \\
RhMnSb & +0.150 & -0.273 & -1.112 & +1.235 \\
IrMnSb & -0.074 & -0.145 & -1.031 & +1.250 
\end{tabular}
\end{ruledtabular}
\end{table}

The total  magnetic moment in $\mu_B$  is just the difference
between the number of spin-up occupied states and the spin-down occupied states. 
In the half-ferromagnetic compounds all spin-down states of the valence band are occupied and thus their 
total number should be, as in a semiconductor, an integer and the total magnetic moment should 
be also an integer since the total valence charge is an integer. A detailed discussion
of the relation between the total moment and the number of electrons will be given in 
Section~\ref{sect5}. Here we notice only that the local moment per unit cell as 
given in Table~\ref{table1} is close to 4 $\mu_B$ in the case of NiMnSb, PdMnSb and
PtMnSb, which is in agreement with the half-ferromagnetic character (or nearly
half-ferromagnetic character in the case of PdMnSb) observed in Fig.~\ref{fig2}. 
Note that due to problems with the $\ell_{max}$ cutoff the KKR method can only give
the correct integer number 4, if Lloyd's formula has been used in the evaluation of 
the integrated density of states, which is not the case in the present calculations.
We also find that the local moment of Mn is not far away from the 4 $\mu_B$ although there
are significant (positive) contributions from the X-atoms and negative 
contribution from the Sb atom. In contrast to this we find that for the 
half-metallic CoMnSb and IrMnSb compounds the total moment is about 3 $\mu_B$. 
Also the local moment of Mn is reduced, but only by about 0.5 $\mu_B$. 
The reduction of the total moment to 3 $\mu_B$ is therefore accompanied by negative Co 
and Ir spin moments, \textit{i.e} these atoms couple antiferromagnetically to the Mn moments.
The reason for this behavior can be understood from the spin-polarized
DOS of NiMnSb and CoMnSb shown in Fig.~\ref{fig3} (middle panel). The
hybridization between Co and Mn is considerably larger than between Ni and Mn. 
Therefore the minority valence band of CoMnSb has  a larger Mn admixture
than the one of NiMnSb whereas the minority conduction band of CoMnSb has a larger Co
admixture than the Ni admixture in the NiMnSb conduction band, while the 
populations of the majority bands are barely changed. As a consequence, the Mn
moment is reduced by the increasing hybridization, while the Co moment becomes negative,
resulting finally in a reduction of the total moment from 4 to 3 $\mu_B$. 
Here we should note that further substitution of Fe for Co leads also to a half-ferromagnetic alloy 
with a total spin magnetic moment of 2 $\mu_B$ as has been already shown by de Groot \textit{et al.} 
in Ref.~\onlinecite{Groot86} and shown by our calculations in Table~\ref{table1}. 
The hybridization in the minority band between the Fe and Mn $d$-states is even larger than between 
the Co and Mn atoms in the CoMnSb compound, resulting in a further decrease of the Mn moment to
around 2.7 $\mu_B$ and a larger negative moment at the Fe site (around $-$0.7 $\mu_B$) 
compared to the Co site,
which stabilize the half-ferromagnetic phase. 
Finally, in the case of RhMnSb the Fermi level is considerably below the gap and thus a part of 
the spin-down states  are unoccupied leading to a total spin magnetic moment 
larger than the 3 $\mu_B$ of CoMnSb and IrMnSb. 

It is also interesting to study the charge transfer in these compounds. In
Table \ref{table2} we have gathered the number of electrons gained (+) or lost(-)
by each atom in the different compounds with respect to the atomic total charge.
To calculate them we simply integrated the total charge density inside each Wigner-Seitz 
polyhedron. It is obvious that the trends in this case are not like in the magnetic 
moments, where we found a similar behavior for isoelectronic systems.
Since the $d$ valence  wavefuctions would have about the same spatial extent for 
neighboring elements in the same row, we expect a similar behavior for compounds 
containing an X atom in the  same row of the periodic table. 
In the case of NiMnSb and CoMnSb we have the largest charge transfer, Ni and Co 
atoms gain $\sim$0.5$e^-$ and Mn and Sb atoms lose around 0.4$e^-$ and 1.4$e^-$,
respectively (results are similar for FeMnSb).  Since the 3$d$ wave functions are less extended than the 4$d$ 
and 5$d$,  the atom-projected charge should be larger for Ni and 
Co than for the other  transition metal atoms; Pd and Rh gain only $\sim$0.2$e^-$ and
for Pt and Ir the charge transfer is very small. 
In all the compounds under study
the vacant site is found to contain a little bit more than one electron. 

\section{Effect of the Lattice Parameter\label{sect4}}

\begin{figure}
\includegraphics[scale=0.42]{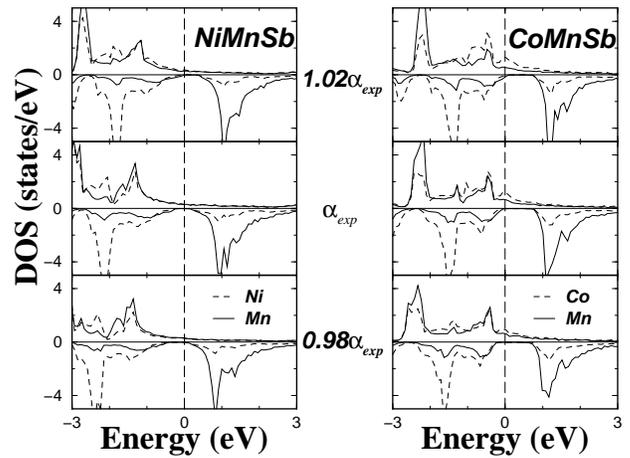}
 \caption{\label{fig3} Atomic- and spin-projected DOS of NiMnSb and CoMnSb when 
we contract (bottom panel) or expand (upper panel) the experimental lattice parameter by 2\%.} 
\end{figure}  

In this Section we will study the influence of the lattice parameter on
the electronic and magnetic properties of the $C1_b$ Heusler alloys.
Before starting this discussion we should mention that the generalized 
gradient approximation (GGA)\cite{GGA} as expected gives better results
than LDA concerning the equilibrium lattice parameters and as it has been 
already shown\cite{iosif} GGA  equilibrium lattice parameter differs only
slightly from the experimental lattice constant. Moreover our  tests showed that 
for the same lattice parameter LDA and GGA give exactly the same results for the gap and
the position of the Fermi level with respect to it, although they produce slightly
different spin magnetic moments. So the use of LDA and of the experimental lattice 
parameter should be considered enough to accurately describe the experimental 
situation.  In Fig. \ref{fig3} we have plotted the DOS for CoMnSb and NiMnSb when
we contract and expand the lattice parameter by 2\% with respect to the 
experimental lattice constant. Either expansion or contraction results in 
a practically rigid shift of the bands with small rearrangements of the shape of the 
peaks to account  for charge neutrality. Expansion moves the Fermi level deeper in energy,
\textit{i.e.} closer to the minority occupied  states,
while contraction moves the Fermi level higher in energy. Especially in the case 
of PdMnSb and IrMnSb, where the Fermi level was near the left edge of the gap, a 
contraction by 2\% is enough to move the Fermi level inside the gap, as it 
was already shown in the 
case of PdMnSb.\cite{iosif} This can be easily understood in terms of the 
behavior of the $p$ electrons of the Sb atom. When we make a contraction
then we squeeze mainly the delocalized $p$ electrons of Sb, as the $d$ electrons of the 
transition metal atoms are already well localized.
So the Sb $p$ states move higher in energy compared to the $d$ electrons of the transition 
metal and the Mn atom and  due to charge neutrality also the Fermi level moves 
higher in energy compared to the experimental lattice constant case. Expansion has
the opposite effect.

\begin{table}
\caption{\label{table3}Spin magnetic moments in $\mu_B$ for NiMnSb and CoMnSb when we 
contract or expand the experimental lattice parameter by 2\%.}
\begin{ruledtabular}
\begin{tabular}{ccrrrrr}
$m^\mathrm{spin}(\mu_B)$ & $a$ ($a_{exp}$) & Ni or Co  & Mn & Sb & Void & Total \\ \hline
NiMnSb & 0.98 & 0.314 & 3.625 & -0.042 & 0.063 & 3.960\\
       & 1.00 & 0.264 & 3.705 & -0.060 & 0.052 & 3.960  \\ 
       & 1.02 & 0.199 & 3.795 & -0.085 & 0.043 & 3.952 \\
CoMnSb & 0.98 & 0.037 & 3.002 & -0.090 & 0.012 & 2.960 \\
       & 1.00 &-0.132 & 3.176 & -0.098 & 0.011 & 2.956 \\
       & 1.02 &-0.305 & 3.371 & -0.102 & 0.013 & 2.976
\end{tabular}
\end{ruledtabular}
\end{table}

Also  the magnetic moments change with the lattice parameter and in Table \ref{table3}
we have gathered, for the  cases shown in Fig. \ref{fig3}, the spin magnetic moments.
As will be discussed below the total magnetic moment does 
not change and remains close to 4 $\mu_B$ for
NiMnSb and to 3 $\mu_B$ for CoMnSb, as in all cases the Fermi level falls
inside the gap and the number of spin-down occupied
states does not change; even in CoMnSb, when we contract by 2\%, the Fermi level is 
just below the left edge of the gap.  
When we expand NiMnSb the Ni spin moment decreases, while the Mn spin moment increases.
Expansion and contraction change the atom-projected charges by  less than 0.01 
electron with respect to the experimental lattice constant.
Thus any moment changes are accompanied by changes of equal magnitude, but 
opposite sign in the population of the local majority and minority states.
When the lattice is expanded, the Mn goes towards a more 
atomiclike situation and its magnetic moment increases while the hybridization 
with the Ni $d$ states decreases.  The opposite situation occurs when we contract 
NiMnSb; Mn moment decreases, hybridization with Ni $d$ states increases and so does the 
Ni spin magnetic moment. A similar trend is observed for CoMnSb. The Mn moments 
increases with expansion and decreases with contraction, while the Co moment,
coupling antiparallel to Mn in the unexpanded lattice, decreases on
expansion to larger negative values and increases on contraction to slightly 
positive values. So finally the lattice parameter
has a significant effect on the magnetic properties of the Heusler alloys, with the local
moments changing substantially, but the total moment and the local charges being 
stable.

\begin{figure}
\includegraphics[scale=0.37]{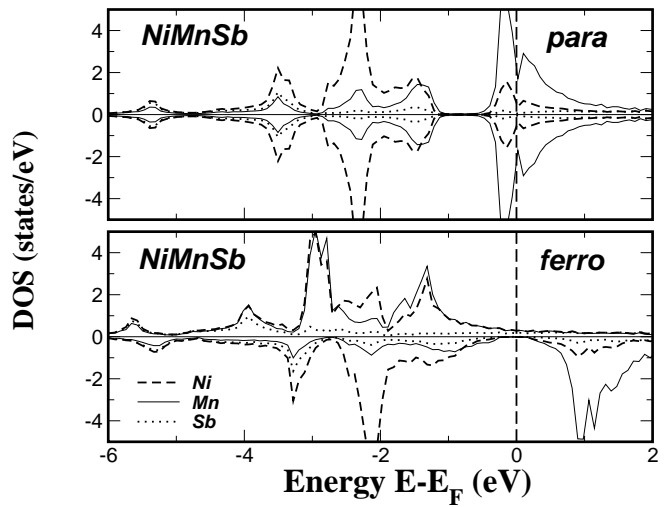}
 \caption{\label{fig4}Ferro- and para-magnetic atomic- and spin-resolved
density of states of NiMnSb. Note that the bonding $d$ states are mainly of 
Ni character, while the antibonding $d$ states are mainly of 
Mn character, and they are separated by a gap.}
\end{figure} 

\section{Origin of the Gap\label{sect5}}

The gap basically arises from the covalent hybridization between the lower-energy
$d$ states of the high-valent transition metal (TM) atom like Ni or Co and 
the higher-energy $d$ states of the lower-valent TM atom, leading to the formation of 
bonding and antibonding bands with a gap in between. The bonding 
hybrids are localized mainly at the high-valent TM atom site while the unoccupied 
antibonding states mainly at the lower-valent TM atom site. For instance in Fig.~\ref{fig4} 
the minority  occupied bonding $d$ states are mainly of Ni character while the 
unoccupied antibonding states are mainly of Mn character. Similarly to the
situation of the elemental and compound semiconductors, these structures 
are particularly stable when only the bonding states are occupied. 
In binary TM alloys this situation usually does not occur, since the total
charge is too large to be accommodated in the bonding hybrids only, or, 
if this is possible, the covalent hybridization is not sufficient to form a gap,
at least not for the close-packed structures of the transition metal alloys (see below).

For these reasons the $sp$-elements like Sb play an important role for the existence 
of the Heusler alloys with a gap at the $E_F$. While an Sb atom has  5 valence 
electrons (5$s^2$, 5$p^3$), in the NiMnSb compound each Sb atom introduces
a deep lying $s$-band, at about -12 eV, and three $p$-bands being pushed 
by hybridization below the center of the 
$d$-bands. These $s$- and $p$-bands accommodate a total of 8 electrons per unit cell, so that 
formally Sb acts as a triple charged Sb$^{-3}$ ion. Analogously, a Te-atom
behaves in these compounds as a Te$^{-2}$ ion and a Sn-atom as a 
Sn$^{-4}$ ion. This does not mean, that locally such a large charge 
transfer exists. In fact, the $p$-states strongly hybridize with the 
TM $d$-states and the charge in these bands is delocalized.  Table~\ref{table2}
shows that locally Sb even looses about one electron. What counts is, that
the $s$- and $p$-bands accommodate 8 electrons per unit cell, thus effectively 
reducing the $d$-charge of the TM atoms. 
Since the bonding $d$-bands introduced above can accommodate 
10 electrons, one expects therefore that the non-magnetic Heusler alloy with 18 
valence electrons per unit cell are particularly stable and have a gap at $E_F$,
i.e. are semiconducting, which requires, of course, a sufficient strong
covalency between the TM partners for the gap to exist. This ``18-electron rule''
was recently derived by Jung \textit{et al.} based on ionic arguments.\cite{jung}
Examples for the semiconducting $C1_b$ Heusler alloys are CoTiSb and 
NiTiSn.\cite{Tobola98}  
In the case of CoTiSb the Sb atom brings 5 valence electrons and the Co and Ti atoms
9 and 4, respectively. Of the 13 TM electrons  3 are catched by the Sb atom, so
that the remaining 10 electrons just fill the bonding $d$-bands. In the case of NiTiSn,
the Ni atom brings in one more electron than Co, but the Sn atom with 4 valence 
electrons catches away in addition 4 $d$-electrons, so that again 10 electrons remain for
 the bonding  $d$ bands. 

Also for systems with more (or less) than 18 electrons, the gap can still
exist. These systems are no longer semiconducting and loose part of the stability,
since then also anti-bonding states are occupied (or not all 
bonding states are occupied). An example is the paramagnetic DOS of NiMnSb,
shown in Fig.~\ref{fig4}. Of the 22 valence electrons, four have to be 
accommodated in the antibonding $d$-bands. The high DOS at $E_F$ signalizes that
the Stoner criterium is met so that in $C1_b$ structure NiMnSb should be 
a ferromagnet. Of the possible magnetic states, the half-metallic states, as shown by 
the spin polarized DOS of NiMnSb in Fig.~\ref{fig4}, is particularly favored
due to the gap at $E_F$ in the minority band. Thus for these half-metallic 
Heusler alloys the 18-electron rule for the semiconducting 
Heusler is replaced by a 9-electron rule for the number of minority electrons.
By denoting the total number of valence electrons by $Z_t$, being
an integer itself,  the
total moment $M_t$ per unit cell is then given by the simple rule 
$ M_t = Z_t -18$ in $\mu_B$, since $Z_t -18$ gives the number of uncompensated spins.
 Thus the total moment  $M_t$ is an integer quantity, assuming the values
0, 1, 2, 3, 4 and 5 if $Z_t \ge$18. The value 0 corresponds to the semiconducting phase 
and the value 5 to the maximal moment when all 10 majority $d$-states are filled.

\begin{figure}
\includegraphics[scale=0.4]{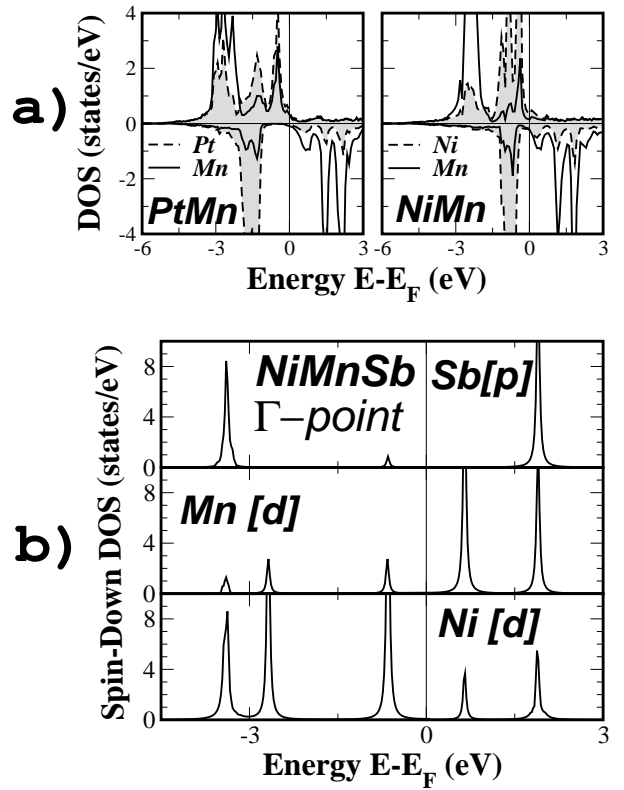}
 \caption{\label{fig7}
$a)$ Atom- and spin-projected DOS for PtMn and NiMn using the
lattice constants of PtMnSb and NiMnSb respectively. The Sb atom 
has been substituted with a vacant site. $b)$ Atom- and
orbital-projected spin-down DOS of NiMnSb projected at the
$\Gamma$ point. We have used a small broadening to calculate the
DOS. }
\end{figure}

Our explanation differs from the one of de Groot and his
collaborators in the pioneering paper of 1983.\cite{groot} There,
the gap has been discussed in terms of the interaction  between
the Mn and Sb atoms, and the Ni atom has been completely omitted
from the discussion. But Mn and Sb are second-neighbors and their
hybridization is much weaker than between the Mn and Ni atoms
which are first-neighbors which clearly shows up in
Fig.~\ref{fig4}. Furthermore the states at the edges of the gap
are of Mn and Ni character, and Sb has a very small weight contrary
to the explanation of de Groot \textit{et al.} that proposed that
the states just below the gap are of Sb character. To further
demonstrate the importance of the $d$-$d$ hybridization for the
gap, we have performed ground state calculations for NiMn in the
zinc-blende structure, \textit{i.e.} by replacing the Sb atom in
the $C1_b$ structure of NiMnSb by a vacant site, but using the
lattice parameter of NiMnSb. The calculations, as shown in Fig.
\ref{fig7}a, yield a pseudogap in the minority gap at $E_F$. For
the hypothetical isoelectronic compound PtMn, the analogous
calculation yields by using the lattice constant of PtMnSb, a real
gap at $E_F$. In this case the minority valence band consists of a
low-lying $s$-band and 5 bonding $d$-bands of mostly Pt character,
while the minority conduction band is formed by 5 antibonding
$d$-bands of mostly Mn character. Thus the rule for the total
moments, $M_t=Z_t-18$, is for this alloy replaced by
$M_t=Z_t-12$, due to the six minority valence bands. For PtMn this
yields $M_t$=17-12=5$\mu_B$, which is indeed obtained in the
calculations. Thus the Sb atom is not
necessary for the formation of the gap  (but
plays a crucial role to stabilize these materials as TM alloys like
PtMn or NiMn do not favor such an open structure).  

To further elucidate the difference between our interpretation and
the one in Ref.~\onlinecite{groot} we have drawn in the lower
panel of Fig. \ref{fig7} the site projected DOS of NiMnSb for the
bands at the $\Gamma$ point; we have used a small imaginary part
of the energy to better see the size of the contributions from
different sites. Note here that the character of the wavefunctions
at the $\Gamma$-point coincides with the character of the
real space wavefunctions.  The three panels show the contribution of the Sb
$p$-states, Ni $d$-states and Mn $d$-states to the bands at
$\Gamma$. We do not present the full band structure since it is
similar to the one already presented in Ref.~\onlinecite{groot}.
There is a band very low in energy, around 12 eV below the Fermi
level, which is provided by the $s$ electrons of Sb (not shown in
the figure). The next triple degenerated band at $\sim$ 3.5 eV
below the Fermi level is originating from the $p$ electrons of Sb
and has a strong admixture of Ni $d$ states. As we have already
mentioned the $p$ states of Sb couple strongly to $d$ TM states
and more precisely to the $t_{2g}$ hybrids which transform
following the same representation in the case of the tetrahedral
symmetry $T_d$. Through this mechanism this band created mainly by
the $p$ electrons of Sb accommodates also transition metal $d$
electrons reducing the effective $d$ charge that can be therefore
accommodated in the higher bands. Just above this band we find the
double degenerated band created by the $e_g$ electrons of the
transition metal atoms. Finally, the triple degenerated band just
below the Fermi level is created by the bonding $t_{2g}$ states of
Ni and Mn with a tiny admixture of Sb $p$ electrons. Above the
band gap we find the antibonding double degenerated $e_g$ TM
states and the triple degenerated $t_{2g}$ states. The band
created by the latter ones has a strong admixture of Sb-$p$
states. This admixture of Sb $p$-states to the Mn-dominated
$t_{2g}$ band is sizeable. However it occurs only at the $\Gamma$
point. When averaged over the full-Brillouin zone, the
Sb-admixture in the conduction band is tiny as can be seen in
Fig.~\ref{fig4}.

Thus our explanation of the valence and conduction bands is
internally consistent. It explains the existence of exactly 9
minority valence bands and simultaneously describes the magnetic
properties of these compounds. With small modifications it can be
also extended to explain the properties of the half-ferromagnetic
full-Heusler alloys like Co$_2$MnGe.\cite{iosifunpubl}
Nevertheless the Sb atoms or in general the $sp$ atoms are
important for the properties of the Heusler alloys. Firstly they
stabilize the $C1_b$ structure, being very unusual for transition
metal compounds. Secondly they provide the four low-lying $s$- and
$p$-bands, which can be filled with additional $d$-states.  Thus by
varying the valence of the $sp$ atoms, the total moments follow
the rule $M_t=Z_t-18$ describing a rich variety of possible
magnetic properties.

Here we want also to point to the similarities of the above moment 
behavior to the well-known Slater-Pauling curve for the TM alloys.\cite{Kubler84} 
The difference is, that in the Heusler alloys the TM minority population is fixed at 
5 and the screening is achieved by filling the majority band, while in
the binary TM alloys the majority band is filled and the charge neutrality is achieved by
filling the minority states. As a result, for the Heusler alloys 
the moment increases with the total charge $Z_t$, while in the 
TM binary alloys it decreases with increasing $Z_t$, since the total
moment is given by $M_t\simeq 10-Z_t$.

\section{Slater-Pauling curve and the role of the lower-valent
transition metal atom \label{sect6}}

\begin{figure}
\includegraphics[scale=0.38]{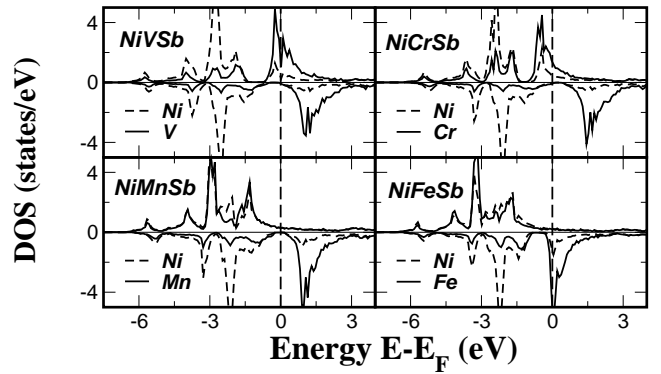}
 \caption{\label{fig5} Calculated atom- and spin-projected density of states of NiYSb, 
with Y= V, Cr, Mn and Fe using the lattice parameter of NiMnSb.
Notice that in the case of the V and Cr compounds also the spin-up band has a gap.} 
\end{figure}

As we mentioned in the last paragraph above the half-ferromagnetic
Heusler alloys follow the Slater-Pauling curve at least when we
change the higher-valent transition element (substituting Co or Fe for Ni).
Here we want to study this behavior, when we vary the valence of the 
lower-valent (\textit{i.e.} magnetic) transition metal atom. 
Thus we  substitute V, Cr and Fe for Mn in the NiMnSb  and CoMnSb 
compounds using the experimental lattice constants of the two Mn compounds. 
Only NiVSb and CoVSb exist experimentally. 
NiVSb crystallizes in a Ni$_2$In hexagonal structure.\cite{NiVSb}  
KKR calculations by Tobola and Pierre  show that this is energetically the most 
favorable structure due to the fact the Fermi level falls inside a peak
of the spin-up states.\cite{Tobola00} On the other hand CoVSb crystallizes in the
$C1_b$ structure presented in Fig.~\ref{fig1} and it has been studied using 
first-principle calculations by Tobola \textit{et al.}\cite{Tobola98} who 
have shown that it is also a half-ferromagnetic material with a total spin moment 
of 1 $\mu_B$. Its experimental lattice constant of 5.801\AA\cite{landolt} is very 
close to the one of CoMnSb (5.853\AA ) and our results are similar for both of them.
  
In Table~\ref{table4} we present the spin magnetic moments for all these compounds
and in Fig.~\ref{fig5} the calculated total DOS for the NiYSb alloys assuming that all
these compounds  are ferromagnetic (results are similar for the CoYSb compounds). 
The compounds containing V and Cr are also half-ferromagnetic as NiMnSb and CoMnSb. 
As can be seen in Table~\ref{table4} when we substitute Cr for Mn we mainly depopulate
one spin-up $d$ state of the lower-valent transition metal atom and thus the 
spin magnetic moment of this atom is reduced practically by 1 $\mu_B$. A similar 
phenomenon occurs when we substitute V for Cr. This behavior is reflected in the smaller
exchange-splitting of the V and Cr $d$ bands compared to the Mn bands as can be seen
in  Fig.~\ref{fig5}. These compounds keep the half-ferromagnetic character 
of NiMnSb and CoMnSb and therefore their total spin magnetic moments follow the 
Slater-Pauling curve imposed by the ``9-electron minority gap'' rule. The compound with 
22 valence electrons (NiMnSb) has a total spin moment of 4 $\mu_B$, the compounds with 
21 valence electrons (NiCrSb and CoMnSb) 3  $\mu_B$, the compounds with 20 valence 
electrons (NiVSb and CoCrSb) 2  $\mu_B$ and finally CoVSb that has only 19 electrons has
a total spin moment of 1  $\mu_B$. Here we should remind that CoTiSb, an 18-electrons
system, is a semiconductor.

\begin{table}
\caption{\label{table4}Spin magnetic moments in $\mu_B$ for the NiYSb and CoYSb compounds
where Y stands for V, Cr, Mn and Fe. For all the calculations we have used 
the experimental lattice constants of the NiMnSb and CoMnSb alloys.}
\begin{ruledtabular}
\begin{tabular}{rrrrrr}
$m^\mathrm{spin}(\mu_B)$ & Ni or Co  & Y & $sp$ atom & Void & Total \\ \hline
NiVSb & 0.139 & 1.769 & -0.040 & 0.073  & 1.941\\
NiCrSb & 0.059 & 2.971 & -0.113 & 0.059 & 2.976  \\
NiMnSb & 0.264 & 3.705 & -0.060 & 0.052 & 3.960 \\
NiFeSb & 0.404 & 2.985 & -0.010 & 0.030 & 3.429 \\
CoVSb  & -0.126& 1.074 & -0.021 & 0.038 & 0.965\\
CoCrSb & -0.324& 2.335 & -0.077 & 0.032 & 1.967  \\
CoMnSb & -0.132& 3.176 & -0.098 & 0.011 & 2.956  \\
CoFeSb &  0.927& 2.919 & -0.039 & 0.012 & 3.819 
\end{tabular}
\end{ruledtabular}
\end{table}

As a last test we have substituted Fe for Mn in CoMnSb and NiMnSb, but both 
CoFeSb and NiFeSb loose their half-ferromagnetic character. In the case of NiFeSb 
the majority $d$-states are already fully occupied as in NiMnSb, thus the additional 
electron has to be screened by the minority $d$-states, so that the Fermi level falls into
the minority Fe states and the half-metallicity is lost; for half-metallicity 
a total moment of 5 $\mu_B$ would be required which is clearly not possible.
For CoFeSb the situation is more delicate.
This system has 22 valence electrons and if it would be a 
half-ferromagnet it should have a total spin-moment of 4 $\mu_B$ like NiMnSb. In reality
our calculations indicate that the Fermi level is slightly above the gap and the total
spin-moment is slightly smaller than 4 $\mu_B$. The Fe atom possesses a comparable 
spin-moment in both NiFeSb and CoFeSb compounds contrary to the 
behavior of the V, Cr and Mn atoms. Furthermore,
contrary to the other Co compounds presented in Table \ref{table4}
in the case of CoFeSb the Co and the lower-valent TM atom (Fe) are ferromagnetically coupled.

\section{Role of the $sp$ Atom\label{sect7}}

\begin{table}
\caption{\label{table5}Spin magnetic moments in $\mu_B$ for the NiMnZ compounds
where Z stands for a $sp$ atom belonging at the 5th row of the periodic table .}
\begin{ruledtabular}
\begin{tabular}{rrrrrr}
$m^\mathrm{spin}(\mu_B)$ & Ni  & Mn & $sp$ atom & Void & Total \\ \hline
NiMnIn & 0.192 & 3.602 & -0.094 & 0.003  & 3.704\\
NiMnSn & 0.047 & 3.361 & -0.148 & -0.004 & 3.256  \\
NiMnSb & 0.264 & 3.705 & -0.060 & 0.052 & 3.960 \\
NiMnTe & 0.467 & 3.996 &  0.101 & 0.091 & 4.656
\end{tabular}
\end{ruledtabular}
\end{table}     

\begin{figure*}
\includegraphics[scale=0.6,angle=270]{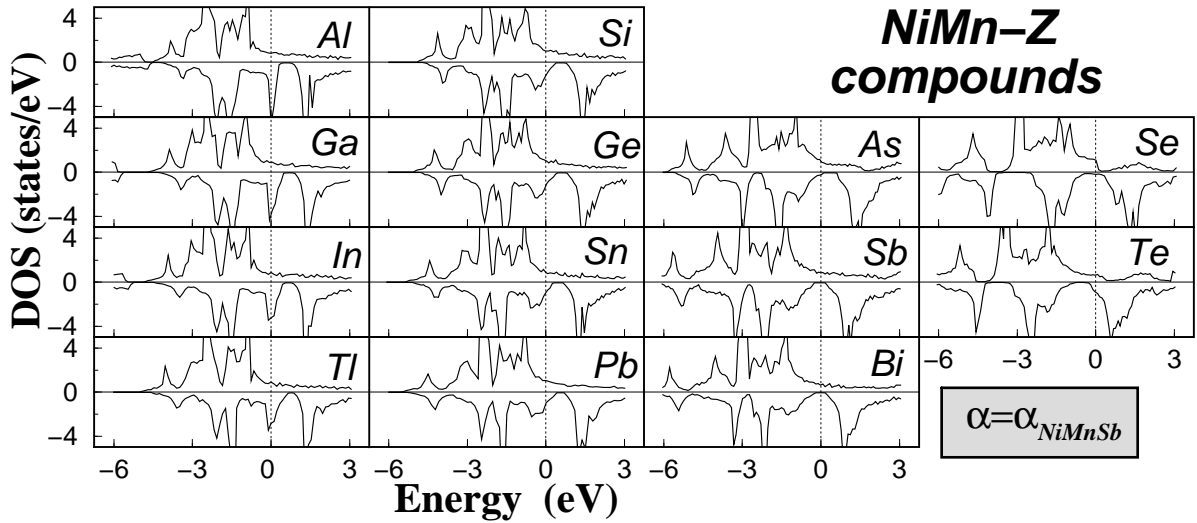}
 \caption{\label{fig6} Spin-resolved density of states of NiMnSb and
 when substituting Sb by another $sp$ atom. All calculations have been performed using
the experimental lattice constant of NiMnSb.}
\end{figure*} 
 
In this Section we  will continue our study on the role of the $sp$ atom in
the electronic properties of a $C1_b$ compound. We performed calculations using
the same lattice parameter as with NiMnSb but substituting Sb by another $sp$ atom
in the periodic table and in Fig. \ref{fig6} we present the obtained spin-projected 
DOS. For the presentation we ordered them following the position of the $sp$ atom
in the periodic table. Firstly we will discuss the trend along a row in the periodic
table and more precisely the substitution of Sb by In and Sn  that have
one and two less valence electrons, respectively, and Te that 
has one electron more than Sb.
It is obvious looking at Fig.~\ref{fig6}
than the substitution of In, Sn or Te for Sb changes the number of valence electrons and
the position of the Fermi level with respect to the gap to account
for this extra or missing $p$ electrons. So for Te the Fermi level is above the
spin-down gap while for Sn it is below the gap and for In it is even lower than Sn. 
The change of the position of the Fermi level is accompanied by changes in the
shape of the peaks to account for charge neutrality.
The same trend can be found along all the other rows presented in this figure.
Thus the change of the valence of the $sp$ atom does not preserve the half-ferromagnetic
character of NiMnSb, but rather leads to a shift of the Fermi level 
analogous to a rigid band model. 
The only exception from this rule is NiMnSe, which is nearly half-metallic. Here we see
big changes in the majority band, where antibonding $p$-$d$ states, which are usually above $E_F$, 
are shifted below the Fermi level, thus increasing the total moment to 4.86 $\mu_B$, \textit{i.e.}
to nearly 5 $\mu_B$.

More interesting is the substitution of Sb by an isoelectronic atom. 
Substituting  Bi for Sb produces a similar  DOS with the Fermi level being  
at the middle of the gap as for NiMnSb. If now we substitute As for Sb 
then the Fermi level moves and now it is found at the left edge of the gap.
In the case of NiMnSb and NiMnBi the majority $p$ states of Sb(Bi) hybridize
stronger with the $d$ states of the TM atoms and thus they extent higher in energy
compared to the $p$ states of As in NiMnAs. As a result the Fermi level is deeper in 
energy in the case of the As-compound. 
Presumably for the correct lattice constant of NiMnAs, being somewhat smaller
than the one of NiMnSb used in the calculation, the Fermi energy would move
again into the gap. 

The spin  moments are similar for compounds containing $sp$ atoms 
in the same column. For example
Ni spin magnetic moment is around $\sim$0.2 $\mu_B$ for the Al, Ga, In and Tl based 
compounds , practically zero for the Si, Ge, Sn and Pb compounds, it increases again
to $\sim$0.2-0.3  $\mu_B$ for the As, Sb and Bi compounds and it reaches $\sim$0.5 $\mu_B$
for the Se and Te compounds. In Table \ref{table5} we present the 
spin magnetic moments for compounds containing In, Sn, Sb and Te to give 
an estimation of the spin moments in the other compounds. 
Of course we should keep in mind that all these
calculations have been performed at the NiMnSb experimental lattice constant and the 
position of the Fermi level would change at the real lattice 
parameters.\cite{Brandao94}
But to our knowledge none of this compounds has been yet 
studied experimentally with the exception of NiMnAs which was found to crystallize 
in the hexagonal Fe$_2$P structure,\cite{NiMnAs} but modern techniques like molecular
beam epitaxy may enable its growth in the $C1_b$ structure. 

In conclusion substituting the $sp$ atom in NiMnSb destroys the half-metallicity. 
In a first approximation, the $sp$-elements act as acceptors and the Fermi level is 
shifted in a rigid band model.

\section{Conclusion\label{sect8}}

In this contribution we have studied in detail the electronic properties 
of the half-ferromagnetic Heusler alloys and mainly focused on the 
appearance of the gap and the related magnetic properties. We have shown using the full-potential version
of the screened KKR method that the gap in the $C1_b$ compounds is imposed 
by the lattice structure. As in the diamond and zinc-blende structures of the 
semiconductors, the hybridization splits both the  spin-up and spin-down $d$ bands.
While systems with 18 valence electrons, like CoTiSb,
are semiconductors, in the systems with a larger number  of valence electrons 
also antibonding states are populated and the paramagnetic phase becomes unstable. 
The half-ferromagnetic phase is 
favorable as the system can gain energy when the Fermi level falls within the
gap of the minority band. It is stabilized by the large exchange splitting 
of the lower-valent transition metal atom and by the $sp$ atom which creates one $s$ 
and three $p$ bands lying low in energy and which accommodate  transition metals 
valence electrons. The minority 
occupied bands  accommodate a total of nine electrons (the minority bonding 
transition-metal $d$ states accommodate 5 electrons) and the total magnetic moment
in $\mu_B$ is just the difference between the total number of valence electrons 
and $2\times 9$. As a result of this simple rule for the total magnetic moment
the half-ferromagnetic Heusler compounds follow the Slater-Pauling curve. 
We have verified this behavior in the  case of the NiYSb and CoYSb compounds where Y accounts 
for V, Cr and Mn. In the case of NiFeSb also minority antibonding states are populated
and the half-ferromagnetic character is lost.
Changing the lattice parameter results in 
shifting the Fermi level without destroying the gap. 
The majority states of the $sp$ atom atom play also an important role as the extent
of the $p$ states through their hybridization with the $d$ states determines 
the position of the Fermi level with respect to the gap. Changing the $sp$ atom results in 
a practically rigid shift of the bands destroying the half-metallicity. 

\begin{acknowledgments}
The authors acknowledge financial support from  
the RT Network of {\em Computational
Magnetoelectronics} (contract RTN1-1999-00145) of the European Commission.
\end{acknowledgments}

\end{document}